\documentclass[letterpaper]{article}
\usepackage{aaai}
\usepackage{times}
\usepackage{helvet}
\usepackage{courier}
\usepackage{graphicx}
\usepackage{amsmath}
\usepackage{booktabs}
\usepackage{multirow}
\usepackage{siunitx}
\usepackage{color}
\begin{document}
%
\title{Quantifying Confounding Bias in Generative Art: A Case Study}
\author{Ramya Srinivasan \and Kanji Uchino\\
Fujitsu Laboratories of America
}
\maketitle
\begin{abstract}
In recent years, AI generated art has become very popular. From generating art works in the style of famous artists like Paul Cezanne and Claude Monet to simulating styles of art movements like Ukiyo-e, a variety of creative applications have been explored using AI. Looking from an art historical perspective, these applications raise some ethical questions. Can AI model artists' styles without stereotyping them? Does AI do justice to the socio-cultural nuances of art movements? In this work, we take a first step towards analyzing these issues. Leveraging directed acyclic graphs to represent potential process of art creation, we propose a simple metric to quantify confounding bias due to the lack of modeling the influence of art movements in learning artists' styles. As a case study, we consider the popular cycleGAN model and analyze confounding bias across various genres. The proposed metric is more effective than state-of-the-art outlier detection method in understanding the influence of art movements in artworks. We hope our work will elucidate important shortcomings of computationally modeling artists' styles and trigger discussions related to accountability of AI generated art.
\end{abstract}
\section{1  Introduction}
From healthcare and finance to judiciary and surveillance, artificial intelligence (AI) is being employed in a wide variety of applications \cite{varun,tom,steven}. AI has also made inroads into creative fields such as music, dance, poetry, storytelling, cooking, and fashion design to name a few \cite{jesse,mariel,liu,varshney,surgan}. Creating portraits, generating paintings in the ``style" of famous artists, style transfer (i.e. transferring the contents of one image according to the style of another image), and creating novel art styles have been some popular applications of AI in art generation \cite{cyclegan,artgan,mazzone,gatys}. 

With the growing adoption of AI, a large body of work has analyzed its ethical impacts in sensitive applications such as in medicine and law enforcement \cite{ziad,joy,kristian,manish}. Of late, there has been considerable interest in understanding AI related biases in creative tasks as well. For example, in \cite{prates}, the authors investigate gender bias in AI generated translations. A recent work by researchers at Allen Institute of Artificial Intelligence demonstrates toxicity in popular language models \cite{kyle}. In \cite{niharika}, the authors show that synthetic images obtained from Generative Adversarial Networks (GANs) exacerbate biases of training data. A notable instance of bias in AI generated art concerns an app called ``AIportraits" that is shown to exhibit racial bias \cite{vice}. It was pointed out that skin color of people of color is lightened in the app's portrait rendition.  

In addition to noticeable biases concerning race, gender, etc., there can be several latent biases in AI generated art, especially in the context of modeling artist's style and style transfer. For example, the authors in \cite{srinivasan} leverage causal models to study several types of biases in modeling art styles and discuss socio-cultural implications of the same. In a similar vein, the authors in \cite{tsila} discuss some of the shortcomings in AI generated art and argue that such art is rife with culturally biased interpretations. 

\subsection{1.1  Motivation} 
Artworks have often been used to document important historical events such as wars, political developments, mythological facts, literary anecdotes, and many aspects of everyday lives of common people \cite{theodore}. For example, ancient Greek art is abundant with mythological paintings depicting Goddesses like Athena and Hera, many Indian artworks portray political and historical events such as the Anglo-Maratha wars and the Anglo-Sikh wars, and ancient Egyptian genre art illustrate culturally rich scenes from lives of ordinary people such as how women prepared food and how people measured harvest. Art movements entail a wealth of information related to culture, politics, and social structure of past times, and these aspects are often not captured in generated art. Furthermore, owing to automation bias exhibited by people \cite{automation}, an AI generated art that fails to justify subtleties of art movements can precipitate bias in understanding history. 

\begin{figure*}[h]
  \centering
    \includegraphics[width= 1\textwidth]{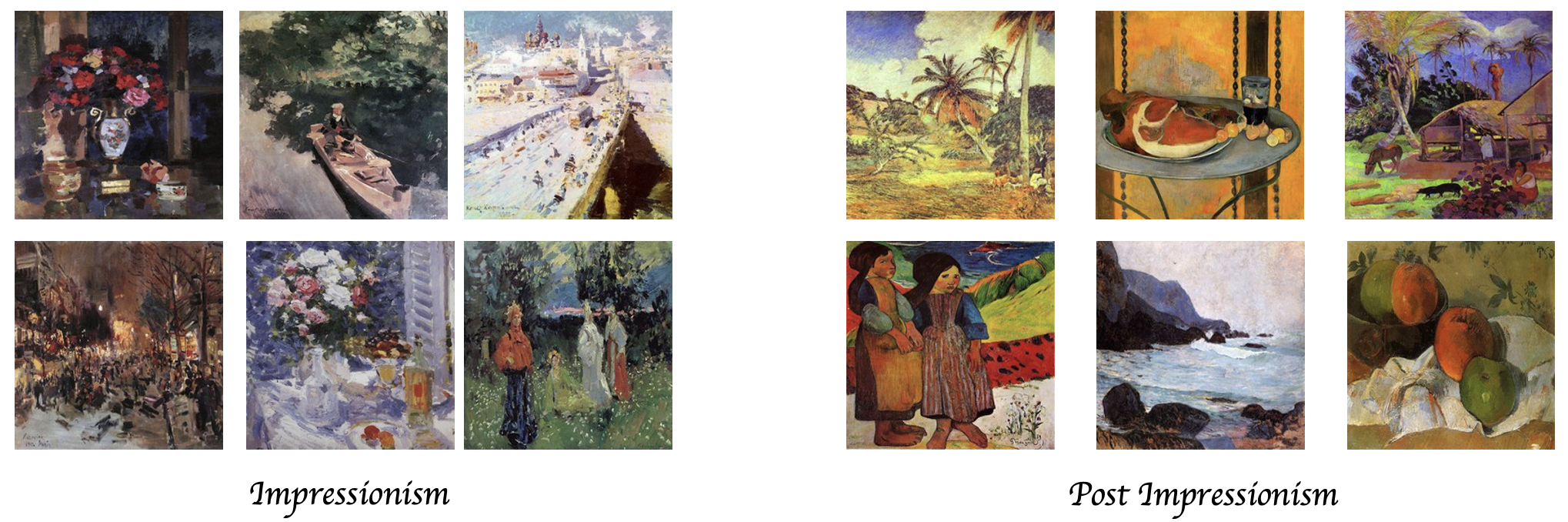}
    \caption{{\small Sample Illustration of Impressionism and Post- Impressionism artworks used in the analysis. Impressionism works were characterized by vibrant colors, spontaneous and accurate rendering of light, color, and atmosphere, focusing mostly on urban lifestyles. Post-Impressionism works focused on lives of ordinary people, depicting emotions and other symbolic contents}}
\end{figure*}
Furthermore, any generated art that claims to mimic artists' styles should not stereotype artists based on a single algorithmically quantifiable metric such as color, brushstrokes, texture, etc. As artist Paul Cezanne describes {\it ``If I were called upon to define briefly the word Art, I should call it the reproduction of what the senses perceive in nature, seen through the veil of the soul"}. Thus, several cognitive aspects such as perception, memory, beliefs, and emotions influence artists and artworks. In reality, many of these aspects can never be observed or measured, and thus the true style of any artist cannot be computationally modeled. Models like \cite{cyclegan} and \cite{artgan} that claim to generate art in the styles of artists like Claude Monet, Vincent Van Gogh, and others are at best capturing correlation features like colors or brushstrokes and overlooking many latent aspects (such as culture and emotions) that characterize artists' styles \cite{hertzman}.

For aforementioned reasons, understanding biases in AI generated art is a necessary task. Given the prevalence of a large number of tools to easily mimic artists ``styles", this task becomes even more pertinent. Prior work has mostly focused on qualitatively analyzing biases in AI generated art \cite{srinivasan,tsila}. In this work, we provide a {\it quantitative} analysis of confounding biases in AI generated art. In general, confounding biases arise due to unmeasured factors that influence both the inputs and outputs of interest. In particular, we quantify the confounding bias due to the lack of modeling of art movement's influence on artists and artworks.

Art movements can be described as tendencies or styles in art with a specific common philosophy influenced by various factors such as cultures, geographies, political-dynastical markers, etc. and followed by a group of artists during a specific period of time \cite{wikiart}. Renaissance art, Modern art, and Ukiyo-e are some examples of art movements. Further, each art movement can have sub-categories. For example, modern art includes many sub-categories such as Dadaism, Impressionism, Post-impressionism, Naturalism, Cubism, Futurism, etc.

Let us consider Impressionism and  Post-impressionism as these are the art movements analyzed in the paper. Figure 1 provides an illustration of artworks belonging to these movements. Although both these movements originated in France, there are marked by subtle differences. Impressionism was characterized by spontaneous brush strokes, vibrant colors, and urban life styles. Impressionists emphasized on accurate depiction of light with its changing quality, precise characterization of movement, and the atmosphere \cite{online}. Post-impressionism originated in reaction to Impressionism. Post Impressionists rejected Impressionists' concern over accurate depiction of color, instead they focused on symbolic depiction of content, formal order, and structure. Post-Impressionism artists focused on lives of ordinary people to naturally depict their emotions and lifestyles \cite{online}. Thus art movement is a dominant factor influencing both the artists and artworks. A model that ignores the influence of art movement in modeling artist's style can thus fail to capture socio-cultural nuances and contribute to confounding bias.

\subsection{1.2  Overview of the Proposed Method} 
As a case study, we consider the cycleGAN model \cite{cyclegan} which has been used to model styles of Paul Cezanne, Claude Monet, and Vincent van Gogh. This is a fully automated method without involving human (i.e. artist) in the loop. Studying biases associated with fully automated AI methods is an essential precursor to understand biases in AI methods that aid artists in completing an art. This is because the latter set of methods can involve both artist and AI related biases, and understanding AI related biases independent of artist specific bias can thus be very beneficial. Therefore, we find the model proposed in \cite{cyclegan} appropriate for our case study.

We consider the influence of Impressionism and Post-Impressionism in modeling artists' styles as these were the dominant art movements that influenced the artists under consideration in \cite{cyclegan}. We evaluate the bias due to lack of consideration of art movement in modeling artists' style in the cycleGAN model across various genres such as landscapes, cityscapes, still life, and flower paintings. It is worth noting that most existing AI methods used to generate art styles largely focus on western art movements.  Ideally, it is important to study the biases in generative art corresponding to non-western art movements, as these art movements are at greater risk being biased due to the already existing social structural disparities. However, due to the paucity of existing AI tools that model multiple art styles of non-western traditions, we have focused on the two aforementioned western art movements for analysis.

Motivated by \cite{srinivasan}, we leverage directed acyclic graphs \cite{pearl} in order to estimate confounding bias. First, causal relationships between art movement, artists, artworks, art material, genre, and other relevant factors are encoded via directed acyclic graphs (DAGs). DAGs serve as accessible visual analysis and interpretation tools for art historians to encode their domain knowledge. As we are interested in understanding the causal influence of the artist on the artwork, in our DAG, artist is the input variable and artwork is the output variable. Art movement, art material, and genres are potential confounders. Next, the minimum adjustment set to remove confounding bias is determined using d-separation rules and backdoor adjustment formula \cite{pearl}. As our goal is to analyze the role of art movement in modeling artists' style, we fix genre and art material across the images used in our analysis. Thus we have to only adjust for art movement. 

The computation of confounding bias is based on the idea of covariate matching \cite{staurt}. Suppose the set of real artworks of an artist $i$ is denoted by $A_i$ and cycleGAN generated images corresponding to the artist is denoted by $G_i$. Further, let $A_j$ where $j \in (1,2, ...n), j \neq i$ be the set of real artworks of other artists belonging to the same art movement as artist $i$. First, a RESNET50 architecture is trained to distinguish between Impressionism and Post Impressionism artworks \cite{he}. Then, using the learned classifier's features representative of the art movement, every element of $A_i$ is matched with its nearest neighbor in $G_i$. Next, every element of $A_i$ is matched with its nearest neighbor in $A_j$. As there can be many artists belonging to the same art movement, we compute nearest neighbor of $A_i$ with respect to all such artists. As all confounders other than art movement are fixed across all the images in the analysis, any difference between the matched pairs should reflect the bias due to lack of modeling art movement. In an ideal scenario where the style of the artist is accurately modeled, the mismatch between $A_i$ and $G_i$ should be low, and the mismatch between $A_i$ and $A_j$ should be high, assuming any two artists have distinct styles of their own. Using these intuitions, we propose a simple metric to quantify confounding bias due to the lack of modeling art movement. We also show how our metric is able to quantify bias that state-of-the-art outlier detection methods \cite{gram} cannot capture.

\subsection{1.3  Insights}
Our findings show that understanding the influence of art movement is essential for learning about artists' style. This is even more important for learning the styles of artists whose works largely belong to one art movement, (e.g. Claude Monet, whose works mostly belong to the Impressionism art movement). This is because the influence of art movement is likely to be higher for such an artist than those whose works span various art movements. We elaborate these insights in Section 6. In reality, the true style of an artist cannot be modeled due to many unobserved confounders such as the emotions, beliefs, and other cognitive abilities of the artist. In this regard, we hope our work triggers inter-disciplinary discussions related to accountability of AI generated art such as the need to understand feasibility of modeling artists' styles, the need for incorporating domain knowledge in AI based art generation, and the socio-cultural consequences of AI generated art. 

The rest of the paper is organized as follows. Section 2 reviews some related work. In Section 3, we provide an overview of directed acyclic graphs that we leverage to model confounding bias. In Section 4, we describe confounding bias with illustrations. In Section 5, we provide an overview of the method. We report results from our experiments in Section 6. We analyze and discuss the implication of the results in Section 7, before concluding in Section 8.
\section{2 Related Works}
There has been a growing interest in using AI to generate art. A good review about AI powered artworks can be found in \cite{miller}. There are a variety of AI models to generate art, generative adversarial networks (GANs) being a prominent type. Models such as \cite{cyclegan}, \cite{mazzone} and \cite{artgan} are just some illustrations of GAN based art generation. In \cite{gatys}, a convolutional neural network architecture is proposed for style transfer. There are also open source platforms that lets end-users to easily create art. For example, \cite{cartoonify} allows a user to convert a photo into a cartoon. With \cite{artbreeder}, one can blend the contents of a photo in the style of another. Platforms like \cite{aiportraits} and \cite{goart} claim to convert a user uploaded photo in the style of famous artists and art movements. 

It is also interesting to note that art has been used to expose bias in the AI pipeline. A very prominent example in this regard is the {\it `Imagenet Roulette'} project by AI researcher Kate Crawford and artist Trevor Paglen \cite{kate}, wherein biases in machine learning datasets are highlighted through art. A convolutional neural network based architecture is proposed in \cite{deepdream} that helps to visualize the workings of various layers in deep networks by creating dream-like appearances. These visualizations can aid in understanding the functioning of various layers. 

Some recent works have exposed biases in AI generated art. For instance, it was reported in \cite{sung,vice} that the AIportraits app \cite{aiportraits} was biased against people of color. In \cite{niharika}, considering synthetic images generated by GAN, the authors point out that GAN architectures are prone to exacerbating biases of training data.  The authors in \cite{tsila} discuss some shortcomings of AI generated art and argue that such art is rife with cultural biases. The closest work to the present work is \cite{srinivasan}, wherein the authors leverage causal graphs to {\it qualitatively} highlight various types of biases in AI generated art. We take a step further: leveraging the model proposed in \cite{srinivasan}, we {\it quantify} confounding bias in AI generated art. This kind of {\it quantitative} analysis provides an objective measure for understanding bias.

\section{3  Directed Acyclic Graphs}
A directed acyclic graph (DAGs) is a directed graph without any loops or cycles. Variables of interest are represented by nodes in the graph and the directed edges between them indicate the causal relations. These directions are often based on assumptions of domain experts and available knowledge. DAGs allow encoding of assumptions about data, model, and analysis, and serve as a tool to test for various biases under such assumptions. DAGs facilitate domain experts such as art historians to encode their assumptions, and hence serve as {\it accessible data visualization and analysis} tools. 

As noted in \cite{srinivasan}, there are several aspects that can characterize an artwork. These include the artist, art material, genre, art movement, etc. The relationships between these various aspects can be determined by domain experts. For example, a domain expert (e.g. art historian) may premise that genre can influence both the artist and the artwork.  DAGs aid in visualizing the relationships between these various aspects. Figure 2 provides an illustration of a DAG encoding one set of such assumptions. It is to be noted that depending on the assumptions of various domain experts, there can be other DAGs describing the relationship of an artwork with the artist, genre, art movement, etc. However, confounding biases can be analyzed separately for each DAG, thereby enhancing the robustness of analysis.

Given a DAG, {\it d-separation} is a criterion for deciding whether a set $X$ of variables is independent of another set $Z$, given a third set $Y$. The idea is to associate ``dependence" with ``connectedness" (i.e., the existence of a connecting path) and ``independence" with ``unconnected-ness" or ``separation" \cite{pearl}. Path here refers to any consecutive sequence of edges, disregarding their direction.

Consider a three vertex graph consisting of vertices $X$, $Y$, and $Z$. There are three basic types of relations using which any pattern of arrows in a DAG can be analyzed, these being as follows. 
\begin{itemize}
    \item $X\rightarrow Y \rightarrow Z $ (causal chain/mediation)
    \item $X\leftarrow Y \rightarrow Z $ (confounder)
    \item $X\rightarrow Y \leftarrow Z $ (collider)
\end{itemize}

In the first case, the effect of $X$ on $Z$ is mediated through $Y$. Conditioning on $Y$, $X$ becomes independent of $Z$ or $Y$ is said to $block$ the path from $X$ to $Z$. 

In the second case, $Y$ is a common cause of $X$ and $Z$. $Y$ is a confounder as it causes spurious correlations between $X$ and $Z$. Conditioning on $Y$, the path from $X$ to $Y$ is blocked. This is the scenario we will analyze in detail in this paper. For example, for the DAG in Figure 2, genre $G$, art movement $A$, and art material $M$ are all confounders in being able to determine the causal effect of artist $X$ on artwork $Z$. The causal effect of artist on artwork captures artist's influence on the artwork, and hence reflective of their style. 

In the last case, $Y$ is a collider as two arrows enter into it. As such, the path from $X$ to $Y$ is blocked. Upon conditioning on $Y$, the path will be unblocked. 
In general, a set $Y$ is admissible (or “sufficient”) for estimating the causal effect of $X$ on $Z$
if the following two conditions hold \cite{pearl}:
\begin{itemize}
{\item No element of $Y$ is a descendant of $X$ }
{\item The elements of $Y$ block all backdoor paths from $X$ to
$Z$—i.e., all paths that end with an arrow pointing to $X$.}
\end{itemize}
Thus we need to block all backdoor paths in order to remove the effect of confounders (which can introduce spurious correlations) in determining the causal effects of interest. With this background, we discuss confounding bias in more detail in the following section.
\section{4  Confounding Bias}

The style of an artist is characterized by several aspects. Some such aspects may be observable (e.g. art material, genre, art movements, etc.) and some others such as emotions, beliefs, prejudices, memory, etc. cannot be perceived or observed. For this reason, the true style of any artist cannot be computationally captured. Our goal is thus {\it not} to computationally model any artist's style, but to analytically highlight the shortcomings in the models that claim to mimic artists' style. As the bias with respect to unobserved cognitive aspects such as emotions, memory, etc. can never be measured, we restrict our analysis to observable aspects. 

\begin{figure}[t]
  \centering
    \includegraphics[width=.5\textwidth]{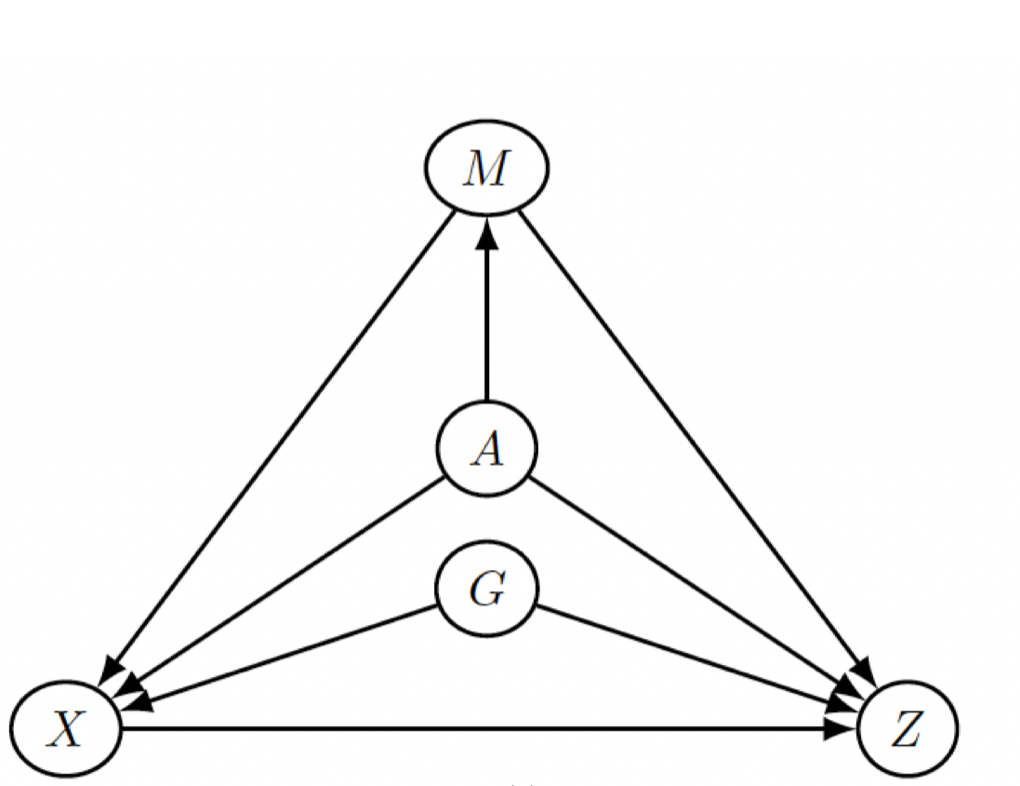}
    \caption{{\small DAG for the case study considered. X: Artist, Z: Artwork, A: Art movement, G: Genre, M: Art material. Image Source: [Srinivasan and Uchino 2020]}}
\end{figure}

We discuss confounding biases that arise due to common causes that affect both the inputs and outputs of interest. In our setting, confounding biases can arise due to factors that affect both artists and artworks. Based on the assumptions encoded in the DAG, such confounders could include art movement, genres, art materials, etc.  A model that does not consider the influence of these confounders is prone to bias. 

For analysis, we consider the DAG provided by \cite{srinivasan} as shown in Figure 2. We will use this as a running example throughout the paper. Here, the variable $X$ denotes the artist, $Z$ denotes the artwork, $G$ is the genre, $M$ is the art material, and $A$ denotes the art movement. In this setting, the problem of modeling artist's style can be viewed as estimating the causal effect of $X$ on $Z$. According to the assumptions encoded in this DAG, art material, genre, and art movement are confounders influencing both the artist and the artwork. Further, art movement influences the art material. Let us assume that all of the confounders are observable. Under these assumptions, in order to compute the causal effect of an artist on the artwork, we have to block the backdoor path from $X$ to $Z$, so as to remove confounding bias. 

In order to block all backdoor paths in Figure 2, one has to adjust for genre, art movement, and art material by conditioning on those variables. The following expression captures the causal effect of $X$ on $Z$ for the graph in Figure 2.
\begin{equation}
CE= \sum_{g,a,m} P(Z|x,G=g,A=a,M=m )
\end{equation}, 
where $CE$ denotes causal effect of $X$ on $Z$.
The summation $\sum_{g,a,m}$ captures the adjustment across all possible art movements, art materials, and genres that the artist has worked, in order to model their style. The implication of finding a sufficient set, $A,G,M$, is that stratifying on $A,G,M$ is guaranteed to remove all confounding bias relative to the causal effect of $X$ on $Z$. 

The above instance depicted a DAG without any unobserved confounders. However, in reality, there are many unobserved confounders such as artist's memory, beliefs, and emotions. In the presence of unobserved confounders, the causal effect of $X$ on $Z$ is not identifiable, implying that the true style of an artist cannot be modeled. The authors in \cite{srinivasan} illustrate this scenario with a DAG as shown in in Figure 3. For the purposes of this work, we will consider only observable confounders and demonstrate the confounding bias that is associated with \cite{cyclegan} in not considering the influence of confounders like art movements to model artists' styles.

\begin{figure}[t]
  \centering
    \includegraphics[width=.5\textwidth]{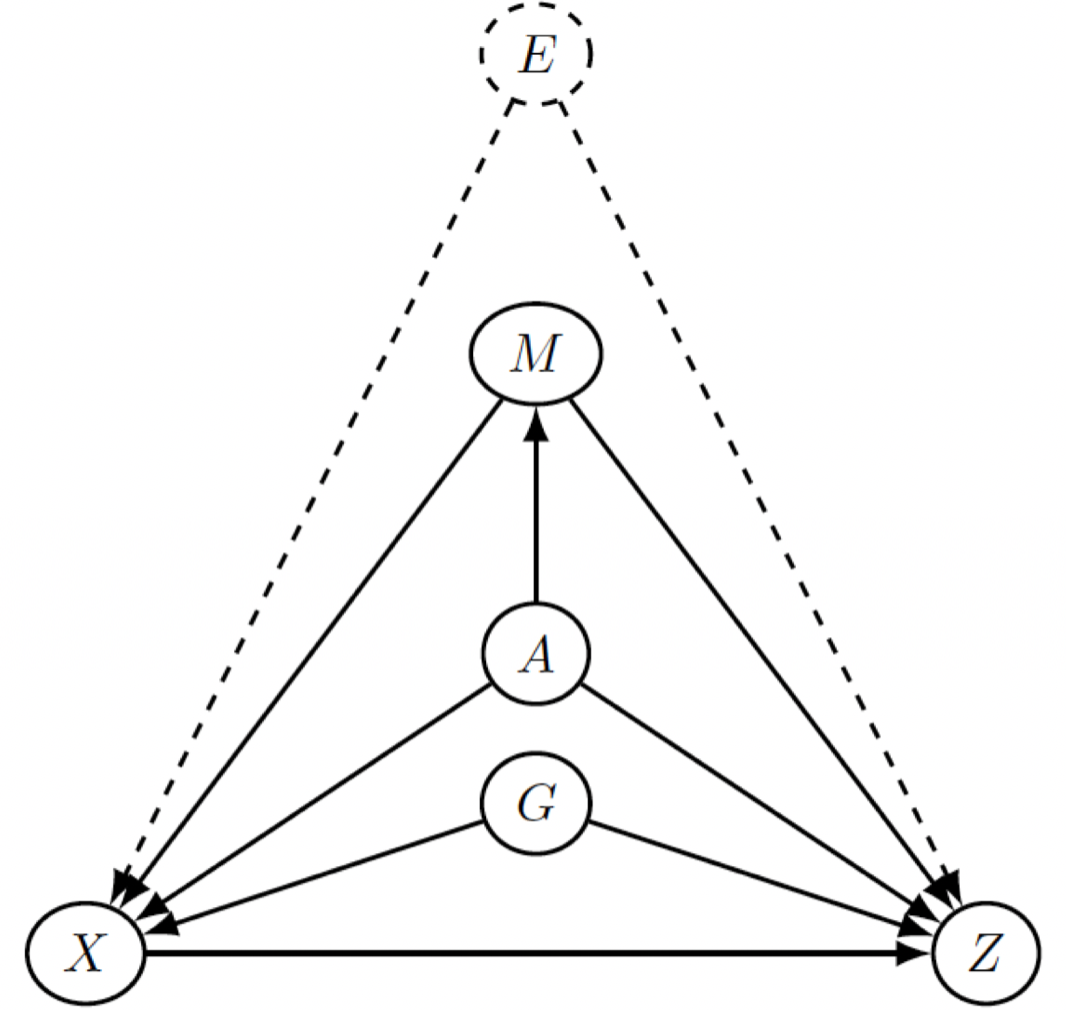}
    \caption{{\small DAG with unobserved confounders. X: Artist, Z: Artwork, A: Art movement, G: Genre, M: Art material, E: Unobserved emotions of the artist. Dotted lines indicate unobserved variables. Image Source: [Srinivasan and Uchino 2020]}}
\end{figure}

Art movements introduced techniques, materials, and themes unique to the culture, society, geographic region, and the times during which these movements gained prominence. Art movements were symbolic of historical, religious, social, and political events of their times. Artists were heavily influenced by the style propagated by the art movement. By not considering the influence of art movement in modeling an artist's style, the social/cultural/religious/political significance associated with the artwork may be lost, and the intent of the artwork may be misrepresented.
In the next section, we describe the proposed method for quantifying confounding bias.

\section{5  Method}
Our goal is to be able to quantify the confounding bias due to the lack of consideration of art movement's influence in modeling artists' styles. Thus, first we need to learn good representations of art movements.
\subsection{5.1 Learning Representations of Art Movements} 
The first step is to learn good representations of the images under study with respect to art movements of interest. We will then use these representations to compute confounding bias (see Section 5.2). We use RESNET50 architecture \cite{he} to learn classifiers for distinguishing Impressionism from Post Impressionism. Then, we extract the learned features from the penultimate layer of the trained network for representing the art movements (please see Section 6.1). In order to learn accurate representations of the art movements under study, we must ensure diversity in the artworks belonging to those art movements, i.e., we must consider artworks across genres and art material belonging to the art movement or else we will be learning a biased representation of the art movement. 

In fact, as part of our experiments, we tried to learn art movements fixing the genre, but this lowered the accuracy of the classifier; thus in order to learn reliable representations of art movements, we need to consider all artworks (across genres, materials, etc.) belonging to the art movement. 
We use RESNET50 \cite{he} to learn features representative of Impressionism and Post Impressionism. Confounding bias in modeling styles of artists Monet, Cezanne, and van Gogh is computed using these learned features across multiple genres such as landscapes, cityscapes, flower paintings, and still life. Next we describe the procedure for computation of confounding bias.
\subsection{5.2 Bias Computation} We fix genre, and art material across all the images considered so that we only have to adjust for art movement as a confounder. However, this does not hurt the generalizability of the method. For multiple confounders, all the elements in the minimum adjustment set have to be adjusted similar to the adjustment of art movement described below.

We leverage the concept of covariate matching in order to adjust for confounders. In our problem setting, we want to be able to estimate the causal effect of an artist say $X=i$, in the presence of a confounder, namely, art movement. Suppose the set of real artworks of the artist is denoted by $A_i$ and the set of generated images of the artist (by the cycleGAN model), is denoted by $G_i$. Specifically, let
\begin{equation}
    A_i=\{a_{i1},a_{i2},...a_{iK}\}
    \end{equation}, where $K$ is the number of real artworks of artist $i$, and let 
   \begin{equation}
    G_i=\{g_{i1},g_{i2},...g_{iL}\}
    \end{equation}, where $L$ is the number of generated artworks of artist $i$, and let \begin{equation}
    A_j=\{a_{j1},a_{j2},...a_{jR}\}
    \end{equation}, denote the $R$ real artworks of an artist $j$ belonging to the same art movement as $i$, and belonging to the same genre as considered in the analysis. Since there can be more than one artist belonging to the same art movement as $i$, assume there are $J$ such artists, so $j \in (1, 2, ..., J), j \neq i $ denotes all these artists. Typically, artists are identified with specific art movements, and such information can obtained from sources like \cite{wikiart}; the set $A_j$ can be constructed using this information. 

First, for each element  $ g_{il}\in G_i$, its nearest neighbor $a_{ilmatch}$ in set $A_i$ is computed based on the values of the confounders, i.e. features representative of the art movement obtained from \cite{he}. Note, that each element in the sets $A_i, G_i, A_j$ is a 1000 dimensional vector. Next, for each element in $a_{ik} \in A_i$, its nearest neighbor $a_{jkmatch}$ in set $A_j$ is computed. As all other potential confounders such as genre and art material are fixed to be the same across all the images considered, the difference in the corresponding matches between sets $A_i$ and $G_i$ is a measure of the variation in (lack of) modeling art movement and in modeling the specific artist's style. Similarly, any difference between the corresponding matches between sets $A_i$ and $A_j$ is a measure of variation across artists' styles and art movements. In an ideal scenario where a generative model is able to accurately learn the style of artist $i$ considering the influence of art movement, the difference between corresponding matches between the real and generated images, i.e., $(A_i-G_i)$ should be close to 0. On the other hand, the difference between matches across artists should be significant compared to the difference between real and generated images of an artist, i.e. $(A_i-A_j) > (A_i-G_i)$, this is because different artists have distinct styles of their own, assuming they do not mimic one another. Using these intuitions, we propose the following metric to quantify confounding bias due to lack of modeling art movement.

\begin{equation}
    bias = \frac {\frac{1}{L}\sum_l  \lvert a_{ilmatch}-g_{il} \rvert }{\frac{1}{J}\sum_j \frac{1}{K}\sum_k \lvert a_{jkmatch}-a_{ik} \rvert }
\end{equation}
\subsection{ 5.2.1 Description of the Metric} The aforementioned metric captures the two intuitions just described. The numerator in the above equation captures the average difference between real artworks and generated images for artist $i$ across all the generated images. The denominator captures the average difference between real artworks of the artist under consideration and other artists belonging to the same art movement. The inner summation and averaging is normalizing with respect to the artist $i$, considering all real artworks of $i$, and the outer summation and averaging is normalization with respect to all $J$ artists belonging to the same art movement as $i$. When the generated images are similar to real artworks of $i$, the numerator is close to 0, this happens when art movement's influence is modeled accurately (amongst other relevant factors) since we consider features representative of art movement in capturing this difference. In a similar vein, the denominator of the above metric will be high when the specific artist's style is learned correctly. So, a low value of the above metric denotes low confounding bias with respect to art movement. Note, the value of the metric can be greater than 1, in which case we assume that there is considerable confounding bias.
\subsection{5.2.2 Choice of Distance Measure} We use Euclidean distance to compute matches. We also tried other distances measures such as Manhattan distance, Chebychev distance, and Wassterstein's distance. Across all distance measures, we observed that the relative order of the bias scores remained the same, thus the metric is not sensitive to changes in the choice of distance measure. 

\section{6  Experiments}
In this section, we report results on computing confounding bias along with an interpretation of the same. We begin by describing experiments on learning representations of art movements.
\subsection{6.1 Learning Representations of Art Movements}
We train a RESNET50 \cite{he} classifier to distinguish between Impressionism and Post-Impressionism, the prominent art movements that were characteristic of artists considered in the cycleGAN model. Specifically, we start with the model pre-trained on Imagenet dataset and fine-tune using the art dataset under study. We then use the learned features from the penultimate layer of the trained model as representations of the art movement, resulting in a 1000 dimensional vector for each image. Note, any state-of-the-art architecture could be used in place of \cite{he}. In order to train the classifier to distinguish between Impressionism and Post Impressionism, we need to consider artworks across artists belonging to those art movements. From \cite{wikiart}, we collected artworks belonging to artists who were identified as belonging to these art movements, and whose majority of the works ($> 50\%$) belonged to Impressionism or Post Impressionism. This ensured collecting artworks representative of the concerned art movements.

We thus collected about 5083 images belonging to Impressionism and about 3495 images belonging to Post Impressionism by crawling images from Wikiart. The dataset consists of Impressionist artists like Berthe Morisot, Edgar Degas, Mary Cassatt, Childe Hassam, Anotonie Blanchard, Claude Monet, Gustave Caillebotte, Sorolla Joaquin, Konstavin Korovin, amongst others. Post Impressionist artists included in the dataset are Vincent van Gogh, Paul Cezanne, Samuel Peploe, Moise Kisling, Ion Pacea, Pyotr Konchalovsky, Maurice Prendergast, Maxime Maufra, etc. Sample illustration of the dataset is provided in Figure 1. We used $80\% $ of the images for training and the rest for validation. We obtained best validation accuracy of $72.1\%$ with Adam optimizer, learning rate = 0.0001, and batch size = 50.

Additionally, we conducted the experiments with other models such as RESNET34, VGG16, and EfficientNet B0-3 to check for any performance improvement. Except for EfficientNet B0-3 being computationally faster, there was not any significant improvement in validation accuracy, so we resorted to RESNET50 features. It is to be noted that Post Impressionism emerged as a reaction to Impressionism. Many artists such as Cezanne worked across these two art movements. Due to these factors, these art movements have subtle differences which are often hard to capture computationally. Quite intuitively therefore, the validation accuracy is not very high. Nevertheless, these learned features serve as a baseline in capturing representations of art movements. Specifically, we used the features from the penultimate layer of the RESNET50. 
\subsection{6.2  Quantifying Confounding Bias across Genres}
We considered various genres such as landscapes, cityscapes, flower paintings, and still life for our analysis. Landscapes depict outdoor sceneries, cityscapes are representations of houses, promenade, and prominent city structures. Flower paintings represent a variety of flowers in vases, gardens, and ponds. Still life consists of images of fruits, vegetables, and other food articles. The DAG in Figure 2 can be used to depict the influence of these genres on the artist and artworks as all the relevant factors are encoded in the DAG. In Section 6, we describe why certain other genres such as portraits cannot be modeled using the DAG shown in Figure 2. 

For the artists under consideration namely, Paul Cezanne, Claude Monet, and Vincent van Gogh, we first obtained real artworks belonging to these genres from the Wikiart dataset. Thus, these images result in three sets $A_i$, where $i \in \{cezanne, gogh, monet\}$ corresponding to the three artists under consideration. In obtaining these images, we fixed the art material to oil painting so that we do not have to adjust for this factor as a confounder. Next, we used random images from existing datasets such as Oxford flower dataset \cite{oxford}, and additionally crawled from Google images to obtain images belonging to various genres under consideration. We then used these images as test images to obtain corresponding generated artworks in the styles of Cezanne, Monet, and van Gogh. These constituted three sets $G_{cezanne},G_{gogh}$ and $G_{monet}$. There were roughly 60 test images in each genre. 

\begin{figure*}[t]
  \centering
    \includegraphics[width=1\textwidth]{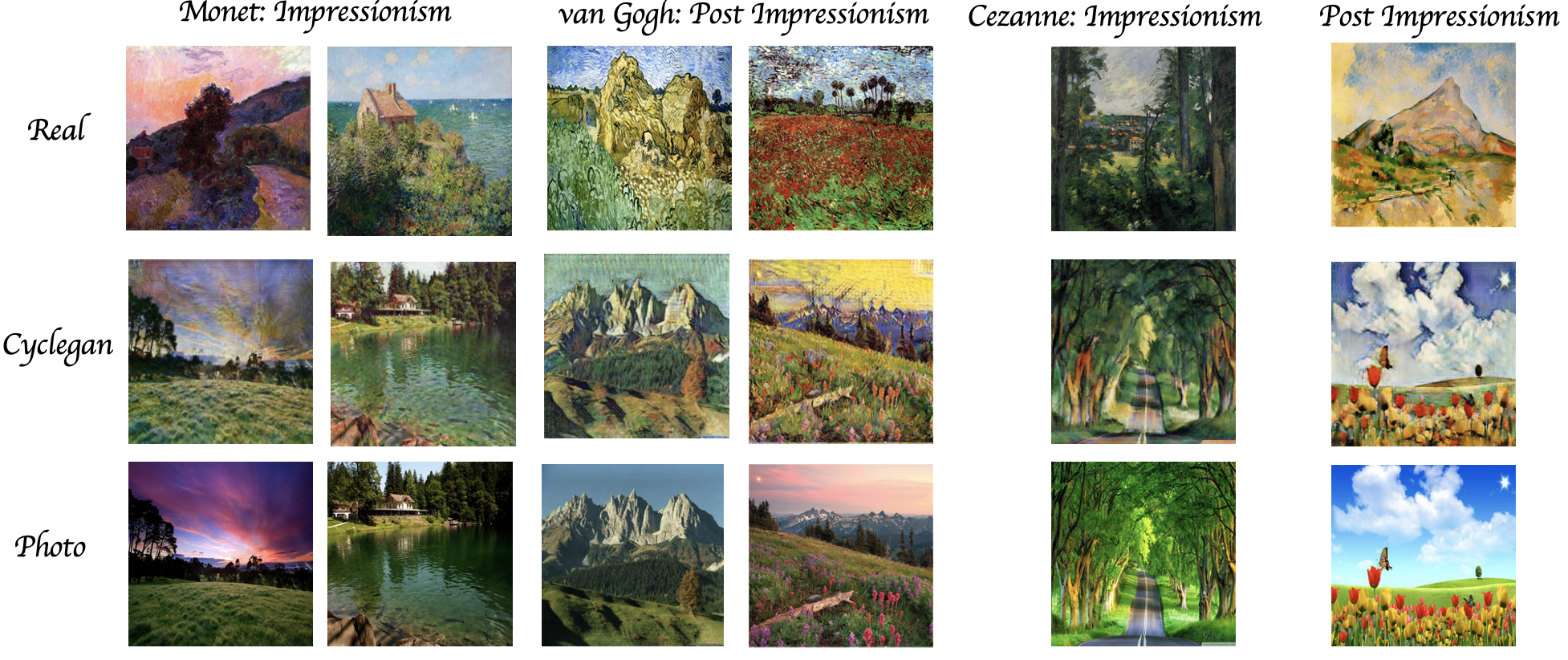}
    \caption{{\small: Top: Real artworks corresponding to the art movements and artists mentioned in the respective columns. Middle: Illustrations of artworks generated by cycleGAN for the corresponding artists and art movements. Bottom: Corresponding photos used to generate images in the middle row. Spontaneous and accurate depiction of light along with its changing quality, an important characteristic of Impressionism is missing in the generated version (row 2, column 1 and 2). Expressive brushstrokes emphasizing geometric forms is missing in generated image corresponding to Post-Impressionist style of van Gogh.}}
\end{figure*}

Next, the sets $A_j$ were constituted using images of other artists who belonged to same art movement as the artist under consideration. As $J$, the number of such artists increases, we can get more reliable indicators of art movements, and thus confounding bias due to art movement will become more evident. It is to be noted that not all artists $J$ necessarily had ample number of images in a particular genre. So, we only considered those artists who had more than 35 images in a particular genre and art movement for analysis within genres. This is because using just a few images of a particular genre by an artist does not help in quantifying bias reliably. For the same reason, confounding bias in modeling artists' styles who had too few images in a particular genre and art movement, cannot be estimated. For example, there are only two landscapes of van Gogh in the Impressionism style, and none for Monet in Post Impressionism, so it is not possible to quantify for confounding bias in landscapes with respect to Impressionism for van Gogh and Post Impressionism for Monet. So, we report results for only those scenarios in which there were at least 35 artworks of the artist in that particular genre.
We then obtained feature descriptors (using the representations from the penultimate layer of the RESNET50 architecture) of the images in sets $A_i,G_i$, and $A_j$ using the learned representations of art movements. Confounding bias was then computed using eq. (5). Table 1 lists the values of this metric for various genres and artists. Blank entries denote cases where there were not ample instances to compute the metric. 
\begin{table}
  \begin{tabular}{lSSSSSS}
    \toprule
    \multirow{2}{*}{G} &
      \multicolumn{2}{c}{Cezanne} &
      \multicolumn{2}{c}{Monet} &
      \multicolumn{2}{c}{van Gogh} \\
      & {Imp} & {Post} & {Imp} & {Post} & {Imp} & {Post} \\
      \midrule
    L & .75 & .78 & 2.52 &  &  & 2.96 \\
    C &  &  & 1.4 &  &  & 1.1 \\
    F &  &   & 1.25 &  &  & 1.34\\
    S& & 1.24 &  &  &  & 1.27 \\
    \bottomrule
    
  \end{tabular}
  \caption{{\small Bias scores computed across genres with respect to Impressionism and Post Impressionism art movements. Blank entries denote cases in which there were not ample instances to compute the metric. Imp: Impressionism, Post: Post Impressionism, G: Genre, L: Landscape, C: Cityscapes, F: Flowers, S: Still life}}
\end{table}


\subsubsection{6.2.1  Observations}
The bias scores are mostly lower for Cezanne who had worked across both Impressionism and Post Impressionism, whereas the scores are higher for van Gogh and Monet who had largely worked in Post Impressionism and Impressionism respectively. This observation suggests that bias scores vary across artists based on the number of art movements influencing them.

To verify, we conducted statistical hypothesis testing. We set the null hypothesis as: the mean of bias scores is same for artists who had worked across art movements and artists who had worked largely in one art movement. Formally, we set the null hypothesis $H_0$ as
\begin{equation}
    H_0: M_s=M_m
\end{equation}, where $M_s$ denotes the mean of the confounding bias scores for artists who largely worked in a single art movement, and $M_m$ denotes the mean of the confounding bias scores for artists who had worked across multiple art movements. The corresponding alternate hypothesis $H_1$ is set as 
\begin{equation}
    H_1: M_s \neq M_m
\end{equation} 
As there were very few observations at our disposal, we used the non-parametric Wilcoxon signed ranked test. The null hypothesis was rejected with a $p$-value of 0.033 ($ \alpha = 0.05$), thus showing that bias scores vary across artists based on the number of art movements influencing them. 

\subsubsection{6.2.2  Interpretation} 
The aforementioned results can be interpreted as follows. If an artist had worked across art movements, then modeling the influence of art movement would be less crucial in generating artworks according to the artist's style. This is because, there are artworks across art movements for such an artist, and thus there is a greater chance of match between generated images and real images due to the greater diversity and variation in the set of real images of the artist. On the contrary, if an artist had worked primarily in one art movement, then it is likely to observe higher bias if the influence of art movement is not considered. This is because the generated images have to match with respect to specific art movement or else they will have greater dissimilarity. 

To elaborate further, let us consider the genre of landscapes. Figure 4 provides an illustration of real landscapes of Monet, van Gogh and Cezanne, cycleGAN generated landscapes in the styles of these artists along with corresponding photos of the generated images. There are about 250 landscapes by Monet in Impressionism style but none corresponding to Post Impressionism. Most of van Gogh's landscapes were set in the Post Impressionism style with just two in the Impressionism style; there are about 35 landscapes by Cezanne in Impressionism and 102 in Post Impressionism.  Consider the photo in row 3 column 1. The corresponding cycleGAN generated image shown in row 2 column 1 does not exhibit the sharp colors of twilight shown in the photo, and alters the affect of the original photo. This is not in line with Impressionism which was characterized by spontaneous and accurate depiction of light with its changing colors. Also, the generated image perhaps does not do justice to the cognitive abilities of the artist; please see image in row 1 column 1 that corresponds to a real landscape by Monet illustrating twilight in the outdoors, with shades of red. In fact, spontaneous and natural rendering of light and color was a distinct feature of Impressionism. In a similar vein, the generated images of van Gogh row 2, column 4 and 5 exhibit markedly different brushstrokes and texture compared to the Post Impressionist works of van Gogh. Post Impressionism works of van Gogh were characterized by swirling brushstrokes, emphasizing geometric forms for an expressive effect.

From Table 1, the bias score with respect to Monet is 2.52 (Impressionism) and 2.96 with respect to van Gogh (Post Impressionism). On the contrary, the bias scores are lower than 1 for Cezanne who had worked across Impressionism and Post Impressionism. Higher scores indicate greater bias thus corroborating with the fact that the bias is higher for artists who were influenced by a single art movement as compared to those who were influenced by multiple art movements.   

\subsection{6.3  Comparison with Outlier Detection Method}

In order to evaluate the effectiveness of the proposed metric, we compared it with a state-of-the-art outlier detection method \cite{gram}. Specifically, the authors in \cite{gram} propose to detect outliers  by identifying inconsistencies between activity patterns of the neural network and predicted class. They characterize activity patterns by Gram matrices and identify anomalies in Gram matrix values by comparing each value with its respective range observed over the training data. The method can be used with any pre-trained softmax classifier. Furthermore, the method neither requires access to outlier data for fine-tuning hyperparameters, nor does it require access for out of distribution for inferring parameters, and hence appropriate for our comparison.

First, we wanted to test if \cite{gram} can detect outliers with respect to real artworks belonging to different art movements. Across all genres, the best detection accuracy of \cite{gram} in identifying outliers with respect to Impressionism (i.e. in separating Post impressionism real artworks from real Impressionism artworks) was just $58.107\%$. As Impressionism and post Impressionism were similar in many aspects, we then tested if \cite{gram} can detect outliers across art movements with marked differences such as in separating Romanticism and Realism from Impressionism and Post Impressionism. Even in this case, the best detection accuracy was $50\%$. Finally, the best detection accuracy of \cite{gram} in separating real artworks from generated artworks was $51.735\%$. Unlike \cite{gram}, the proposed metric is more effective in capturing the influence of art movements in modeling artists' styles since the bias scores corresponding to artists who had largely worked in a single art movement is significantly higher than those who had worked across multiple art movements. In the next section, we also discuss the other benefits of the proposed bias metric.
\section{7  Discussion}
In this section, we discuss a few other relevant questions in the context of the above results. 
\subsection{7.1 What happens if images across art movements are combined in analyzing confounding bias?}

The very goal of estimating confounding bias is to be able to capture the drawbacks due to lack of modeling art movements. When images across art movements are combined, the fact that art movement is a potential confounder is ignored, thereby leading to biased representations. Thus, confounding bias has to be computed with respect to Impressionism and Post Impressionism separately. Computing bias across a combination of images from these two art movements is an illustration of {\it ``Simpson's paradox"} \cite{dana}. 

Simpson's paradox is a trend that characterizes the inconsistencies across different groups of the data. Specifically, an effect that appears across different sub groups of data but that which gets reversed when the groups are combined illustrates Simpson's paradox. In other words, Simpson paradox refers to the effect that occurs when association between two variables is different from the association between the same two variable after controlling for other variables. The correct result ( i.e. whether to consider aggregated data or data corresponding to sub-groups) is dependent on the causal graph characterizing the problem and data. 

The authors in \cite{dana} illustrate Simpson's paradox with several real examples. For example, the authors cite a study of thyroid disease published in 1995 where smokers had a higher survival rate than non-smokers. However, the non-smokers had a higher survival rate in six out of the seven age groups considered, and the difference was minimal in the seventh. Age was a confounder of smoking and survival, and hence it had to be adjusted for. The correct result corresponds to the one obtained after stratifying data by age, and thus it was concluded that smoking had a negative impact on survival.

Let us revisit our example. According to the DAG in Figure 2, art movement is a confounder which needs to be adjusted for. If however, we overlook this confounder by combining images across art movements, then the confounder is not adjusted according to eq. (1). In fact, when we combined images across art movements, the resulting bias score was lower, however this result is incorrect, thus illustrating the paradox. 

In cycleGAN \cite{cyclegan}, the authors propose a cycle consistency loss such that the generated images when mapped back to the original (real) images are indistinguishable from the original images. This in turn implies that the generated images are as realistic as possible.  

Simpson's paradox elucidates why cycleGAN that is trained on data combined across art movements and whose loss function intuitively appears sound, cannot capture the influence of art movements. Because the loss is being minimized across images from different art movements, it is not guaranteed to minimize the loss within each art movement. Results in Table 1 and Figure 4 illustrate this point further. Thus, in order to accurately model artists' styles, \cite{cyclegan} had to minimize the loss proposed by stratifying the data by art movement. 

\subsection {7.2 What about other genres such as portraits or genre art?}

The computation of bias is based on the DAG provided. The DAG considered in the case study is not applicable to other genres such as portraits and genre art. Portraits, for example, involve many other factors in their creation. Characteristics such as gender, age, beauty, and other aesthetics play a prominent role in the way sitters are depicted. Also, factors characterizing sitter's lineage/genealogy (e.g. race, family, cultural background, religion, etc.) can also influence the rendition. The social standing of the sitter such as their profession, political backgrounds, and power could influence the artists in the way they depict the sitter. For example, it is possible that powerful people commanded the artists to depict them in a certain way, and the artists thereby had to exaggerate certain characteristics.

Genre art depicted everyday aspects of ordinary people. These artworks encompassed a variety of socio-cultural themes such as cooking, harvesting, dancing, etc. Therefore genre art involves many socio-cultural factors that the DAG considered in the case study does not entail. Thus, for computation of bias in other genres, appropriate DAGs have to be constructed in consultation with art historians, taking into account all relevant variables of interest.

\subsection{7.3 What are some potential applications of the proposed metric? }
As discussed in the previous sections, the proposed metric is useful in quantifying the confounding bias associated with generative AI methods that fail to consider  art movements and other confounders in modeling artists' styles. 

Such an objective assessment of bias can also be useful in authenticating artworks, i.e. the computed bias scores can aid in verifying if an artwork was a genuine creation of a particular artist. This is because, if an artwork is not a real work of an artist, then the bias score associated with such a work is likely to be higher, and similarly, if the artwork is a genuine creation of the artist, then the bias score is likely to be lower. It is to be noted that we are {\it not} claiming that the bias score alone is sufficient to validate the authenticity of an artwork; instead, we believe it can be beneficial in assessing the authenticity of artworks along with other forms of evidences, including those of art historians. A related application of the proposed metric would be for price assessment of generative art. In other words, the computed bias scores can serve as a measure of the selling price/value of a generative artwork. If the bias score is high, then the value of generative artwork is likely to be low, and vice versa. 

Finally, the proposed metric can also aid in the study of art history. The computed bias scores can provide an independent and complementary source of evidence to art historians to verify their assumptions or opinions regarding various topics of interest such as in understanding characteristics of art movements, and in studying influence of specific art materials on artists. By considering different DAGs that encode assumptions of different art historians, it is also possible to compare perspectives and understand if there are sources of bias that are common across assumptions of different art historians. Such common bias sources will then serve as a strong evidence for art historians in accepting or rejecting a viewpoint. 
\section{8 Conclusions}
Art movements influenced the style of artists in many subtle ways. Overlooking the contribution of art movements in modeling artists' styles leads to confounding bias. In reality, there are several unobserved factors such as emotions, and beliefs that characterize an artist's style. Thus, it is not possible to computationally model an artist's style. In doing so, generative art might be stereotyping artists based on a narrow metric such as color or brush strokes, and not do justice to the artist's abilities. Furthermore, generated artworks might accentuate automation bias by conveying inaccurate information about socio-political-cultural aspects due to their inability in capturing the nuances depicted in art movements. In this work, leveraging directed acyclic graphs, we proposed a simple metric to quantify confounding bias due to the lack of modeling art movement's influence in learning artists' styles. We analyzed this confounding bias across genres for artists considered in the cycleGAN model, and provided an intuitive interpretation of the bias scores. We hope our work triggers discussions related to feasibility of modeling artists' styles, and more broadly raises issues related to accountability of AI-simulated artists' styles.

\bibliographystyle{aaai}
\bibliography{aies}
\end{document}